# On the incomplete recurrence of modulationally unstable deep-water surface gravity waves

Alexey Slunyaev[1,2)*] and Alexander Dosaev[1)]

1) Institute of Applied Physics, Nizhny Novgorod, Russia
2) Nizhny Novgorod State Technical University, N. Novgorod, Russia

* Slunyaev@appl.sci-nnov.ru

**Abstract**

The issue of a recurrence of the modulationally unstable water wave trains within the framework of the fully nonlinear potential Euler equations is addressed. It is examined, in particular, if a modulation which appears from nowhere (i.e., is infinitesimal initially) and generates a rogue wave which then disappears with no trace. If so, this wave solution would be a breather solution of the primitive hydrodynamic equations. It is shown with the help of the fully nonlinear numerical simulation that when a rogue wave occurs from a uniform Stokes wave train, it excites other waves which have different lengths, what prevents the complete recurrence and, eventually, results in a quasi-periodic breathing of the wave envelope. Meanwhile the discovered effects are rather small in magnitude, and the period of the modulation breathing may be thousands of the dominant wave periods. Thus, the obtained solution may be called a quasi-breather of the Euler equations.

## 1. Introduction

The Fermi-Pasta-Ulam recurrence is a fascinating phenomenon of the nonlinear dynamics, which attracts much scientific interest (see, e.g. Chaos Focus Issue [1]). The recurrent dynamics has been reported in systems of various physical contexts, e.g. in plasmas, optics, hydrodynamics, etc., which in the first approximation may be described by integrable nonlinear evolution equations ([2-4] and many others). The nonlinear Schrodinger equation (NLS) is a familiar example which exhibits the recurrent wave dynamics driven by the modulation instability (or Benjamin – Feir or side-band instability, see [5-9]). The problem of a long-term recurrence has been further developed in the recent years thanks to the advanced numerical simulations, which help to efficiently analyze complicated initial-value problems and also non-integrable systems [10-12].

Recently the modulational instability has received interest in the context of sea wave dynamics due to the so-called rogue waves, which are enormously large oceanic waves which are likely to occur too frequently than could be expected [13, 14]. These waves "appear from nowhere and disappear without a trace" [15]; this feature is in common with the breather solutions of the nonlinear Schrodinger equation, as was pointed out in [16, 17]. Breathers of the NLS equation are strongly coherent groups which correspond to the nonlinear superposition of an envelope soliton and a uniform background wave (see, e.g., [16]). Today the NLS breathers are the most popular academic prototype for the rogue wave dynamics.

The NLS equation is known to be a rather crude model for water waves; its derivation implies the assumptions of small wave steepness and slow modulations, which are frequently broken in the realistic sea conditions. The issue of existence of breather-like wave patterns beyond the limitations of the idealized NLS theory may be related to the question of deterministic predictability of the rogue waves, which would be of great practical value. In a more academic context, the case of existence of soliton-like wave structures (including breathers) might favor the conjecture of integrability in some sense of the potential water equations. These reasons explain the considerable interest in reproducing the modulational instability and NLS breathers in the laboratory conditions (e.g., [18-21]), and also the dedicated numerical simulations ([17, 22-24] and many others). In all the cited references the significant wave enhancement caused by the unstable dynamics was followed by the saturation (if the waves did not break in the focus). However, the recurrence was always *approximate* in contrast to the exact breather solution of the NLS equation. The effect of high-order terms in the NLS equation on the recurrence process was studied in a recent paper [25] (see also references on the previous research therein).

A number of obvious reasons could be suggested, why the modulation evolution observed in the laboratory and the computer simulations was not fully recurrent (see the next section). Meanwhile the following fundamental question may be posed: Do the primitive equations of hydrodynamics possess a breather solution? In other words, do the wave trains exist which generate rogue waves appearing from nowhere and disappearing without a trace – in the strict sense, having the feature of a homoclinic orbit? The formulated question is much idealized but, as discussed above, is of considerable practical interest; it is addressed in this paper by means of advanced numerical simulations.

In the following Sec. 2 we introduce the exact breather solution of the NLS equation, and discuss in Sec. 3 spurious effects which can impede its complete recurrence in the simulation by computer; there we also formulate in more details the particular problem for the present study to be tackled. In Sec. 4 we study the recurrence of numerically simulated NLS breathers with the purpose to observe the limits of accuracy of the employed approach and the appearance of the computer roundoff error. In Sec. 5 the results of the fully nonlinear numerical simulations of the potential Euler equations are reported. In Sec. 6 we discuss the physical effects which underlie the incomplete recurrence observed in the presented numerical simulations. The main conclusions follow in the end.

## 2. The breather solution of the NLS equation: a summary

In the deep-water limit the nonlinear Schrodinger equation for unidirectional gravity wave modulations reads

$$i\left(\frac{\partial A}{\partial t} + C_{gr}\frac{\partial A}{\partial x}\right) + \frac{\omega_0}{8k_0^2}\frac{\partial^2 A}{\partial x^2} + \frac{\omega_0 k_0^2}{2}|A|^2 A = 0, \qquad (1)$$

where $A(x, t)$ is the complex function, which describes the leading-order approximations for the surface displacement, $\eta(x, t)$, and the surface velocity potential, $\Phi(x, t)$,

$$\eta = \text{Re}\, A e^{i\omega_0 t - i k_0 x}, \qquad \Phi = -\frac{\omega_0}{k_0}\text{Im}\left(A e^{i\omega_0 t - i k_0 x}\right). \qquad (2)$$

Real parameters $k_0$ and $\omega_0$ are the wavenumber and the frequency of the carrier related by the dispersive relation $\omega_0^2 = gk_0$, where $g$ is the gravity acceleration, and $C_{gr} = \omega_0/k_0/2$ is the group velocity.

It is convenient to apply the following change of variables

$$x' = \sqrt{2}s_0 k_0(x - C_g t), \qquad t' = s^2 \frac{\omega_0}{4} t, \qquad q = \frac{k_0}{s_0}\eta, \qquad (3)$$

which reduces (1) to the dimensionless form of the NLS equation

$$i\frac{\partial q}{\partial t'} + \frac{\partial^2 q}{\partial x'^2} + 2|q|^2 q = 0.\qquad(4)$$

It is also convenient to equate the free parameter $s$ in (3) with the steepness $\varepsilon$ of the uniform wave with amplitude $a_0$, $s = \varepsilon$, where $\varepsilon = k_0 a_0$. Then the function $q(x)$ will have the unit amplitude regardless the magnitude of $a_0$. The periodic in space breather solution (so-called Akhmediev breather, [26]) for Eq. (4) has the form

$$q(x,t) = e^{2it+i\varphi_0}\frac{\lambda\cos[2\gamma(x-x_0)] - \cosh(4\lambda\gamma t - 2i\mu)}{\lambda\cos[2\gamma(x-x_0)] - \cosh 4\lambda\gamma t},\quad \gamma = \sqrt{1-\lambda^2},\quad \lambda = \cos\mu,\qquad(5)$$

where $\mu \in [0, \pi/2]$ (or the corresponding $0 < \lambda \le 1$) is the main free parameter; $x_0$ and $\varphi_0$ are the arbitrary initial location of the breather and the phase correspondingly. The primes over $x$ and $t$ are omitted hereafter. The solution (5) possesses the spatial period $L = \pi/\gamma$. When $t \to \pm\infty$ the modulation attenuates, and $|q| \to 1$ (see more details in [24]). It is also clear from (5) that the modulation $|q|$ is invariant with respect to the time inversion.

Periodic breathers (5) are most appropriate for the numerical simulations by the algorithms using the discrete Fourier transform. The Fourier transform for (5) reads

$$\hat{q}_k = e^{2it}\left[M - 2\pi B_k e^{i\theta}\right],\quad M = \frac{\pi}{\gamma}\delta_k,\quad \theta = \arctan\left(\frac{\lambda}{\gamma}\tanh 4\lambda\gamma t\right),\qquad(6)$$

$$B_{k=0} = 1,\quad B_{k\ne 0} = \left(p - \sqrt{p^2-1}\right)^{\frac{|k|}{2\gamma}},\quad p = \frac{1}{\lambda}\cosh 4\lambda\gamma t.$$

Here real numbers $k$ denote the discrete wavenumbers, so that $k/(2\gamma)$ are integers counting the spectral satellites; $\delta_k$ – is the Kronecker delta. Note that the breather is represented by an infinite number of sidebands $B_k$; the tails of the Fourier transform decay exponentially. According to (6) the phase $\theta$ varies in time; however a noticeable variation of $\theta$ occurs only in the vicinity of the focusing point, when it changes the sign. In the limit $t \to \pm\infty$ the phase tends to $\theta_{\pm\infty} = \pm\mathrm{atan}(\lambda/\gamma)$, thus the modulation-demodulation process changes the wave phase; its magnitude depends on the parameter $\lambda$. The most unstable modulation (see (9), (10) below) corresponds to $\lambda = 2^{-1/2}$, and then $\theta_{\pm\infty} = \pm\pi/4$. Note that all the Fourier harmonics which describe the modulation, $B_k$, are characterized by the same phase $\theta$.

The most energetic sidebands $B_k$ in (6) correspond to the longest perturbation; they are characterized by the wavenumber offset $\pm\Delta k$,

$$\Delta k = \frac{2\pi}{L} = 2\gamma.\qquad(7)$$

In the course of the modulation growth the sidebands $B_{k\ne 0}$ receive the energy from the carrier $k = 0$. After the event of focusing, the energy is transferred back from the sidebands, hence the solution (5), (6) describes the complete recurrence of the envelope shape.

According to the classic analysis, the modulational instability of a plane wave with a given amplitude occurs when the modulation is long enough. The deterministic counterpart of the Benjamin – Feir Index, $R_{BFI}$, may be introduced,

$$R_{BFI} = 2\sqrt{2}\varepsilon N_x = \frac{\sqrt{2}}{\pi}\varepsilon k_0 L,\qquad(8)$$

then the wave is modulationally unstable if $R_{BFI} > 1$. In general, $M$ unstable modes may exist where $M$ in the integer part of $R_{BFI}$. In (8) $N_x = k_0 L/(2\pi)$ denotes the number of carrier wave lengths per one modulation length (the latter coincides with $L$ in what follows), and $\varepsilon$ is the steepness of the unperturbed wave. The modulation growth rate $\sigma_{BF}$ is given by the following expression,

$$\frac{\sigma_{BF}}{\omega_0} = \frac{\varepsilon^2}{R_{BFI}} \sqrt{1 - \frac{1}{R_{BFI}^2}}. \qquad (9)$$

Most importantly, the parameter of the breather solution (5) $\lambda$ and the parameter $R_{BFI}$ are related through

$$\lambda = \sqrt{1 - \frac{1}{R_{BFI}^2}}, \qquad (10)$$

and then $\gamma = 1/R_{BFI}$. Thus, within the weakly nonlinear theory for weakly modulated waves breathers exist (i.e., $\lambda$ is real) when the modulation is unstable ($\sigma_{BF}$ is real) and vice versa. Breathers describe the nonlinear stage of the modulational instability. Longer perturbations cause gathering of the wave energy from a larger area, what results in a stronger wave amplification,

$$\max|q| = 1 + 2\lambda, \qquad (11)$$

in a smaller spatial interval, see more details in [24].

## 3. The circumstances which can prevent the full recurrence of the modulational instability

The fact of incomplete recurrence of the NLS breathers in laboratory flumes was particularly emphasized in [21]. After the modulationally unstable wave group reaches its maximum, it attenuates but does not precisely recover the shape, which it had before the focusing. In particular, the magnitude of the modulation does not diminish to the level as small as of the initial condition. It is reasonable to assume that the imperfect laboratory conditions could disturb the dynamics and affect the wave dynamics. Such perturbations could be insufficiently accurate wave train generation by the paddle; dissipation due to the viscosity, surface tension, bottom friction and near-wall effects; 3D effects, etc. Thus, unavoidable *perturbations* could be the reason why a breather does not recover.

On the other hand, a qualitatively similar observation of incomplete recurrence follows from the series of numerical experiments [21], where breathers were simulated by means of the fully nonlinear potential Euler equations formulated in the conformal variables. In those simulations, only the effects of inappropriate initial condition or inaccurate numerical integration could affect the breather recurrence. As follows from (6), a breather is represented by the coherent dynamics of many sidebands in the Fourier space. As the exact breather-type solution of the Euler equations is not known (if it exists at all), the initial conditions in [21] were based on an approximate approach, thus could contain unwanted perturbations. The erroneous excitation of the Fourier harmonics may cause the exponential growth of the perturbations with different unstable wavelengths. As a result, *other unstable modes* may grow simultaneously with the 'wanted' breather and thus can strongly disturb the solution. This dynamics was first observed and explained in the framework of the NLS equation in [7]. The perturbations may be a numerical noise which cannot be removed completely, thus the situation when the excitation of one unstable mode leads to the development of all available unstable modes represents the general picture (see, e.g., in [12, 27]). The impact of different numerical effects on the accuracy of the simulation of the Akhmediev breathers was also in the focus of the study [28].

In the present paper we aim at controlling the unwanted numerical effects caused by the discretization in space and time, by round-offs, and the associated spurious numerical artifacts. The single-mode regime of the modulational instability is imposed by the choice of a sufficiently short perturbation, $L$, in the periodic spatial domain of the same size, so that only one Fourier mode satisfies the instability condition. The mentioned above problem of the unknown proper initial condition is bypassed by the use of a very small initial perturbation. In our setting the inoculating modulation of the uniform wave is introduced at the two adjacent wavenumbers of the discrete Fourier grid, $B_{\pm 1}$, with the magnitude close to the level of the computer precision. The magnitudes of the other Fourier modes (6), $B_{|k|>1}$, are below the data precision.

In what follows, *a breather solution* implies one mode of the modulational instability of the Stokes wave which starts to grow from an infinitively small perturbation, focuses to a large wave group and then returns back to the state of a uniform Stokes wave with an infinitesimal perturbation. It should appear from nowhere and disappear without a trace in full accordance with the rogue wave attribute suggested in [15].

The instability of steep waves develops faster (see (9)) and, roughly speaking, results in steeper wave events, which are the most interesting case. Meanwhile only non-overturning waves may be simulated by the code of the potential Euler equations. For the given wave steepness $\varepsilon$ and dominant wavenumber $k_0$ the length of the computation domain, $L$, should be large enough to favor the modulational instability, while the twice shorter modulation should be stable to provide the single-mode regime. Thus within the limits of the NLS theory the following range of the parameter $R_{BFI}$ should be considered, $1 < R_{BFI} < 2$, according to (8). Accounting for the strongly nonlinear effects, the results of the numerical simulations [24], were used. The following conditions are considered in detail in this paper: $\varepsilon = 0.11$ and $N_x = 4$ (i.e., the modulation length is equal to four dominant wave lengths), which corresponds to $R_{BFI} = 1.24$. Simulations of a few other conditions are briefly reported in Slunyaev (2017).

## 4. Simulations of the nonlinear Schrodinger equation

In this section we describe the results of auxiliary numerical simulations of the nonlinear Schrodinger equation. They support the main statements of the discussion above, and help to estimate the significance and appearance of possible spurious numerical effects. Overall, four pseudo-spectral codes for the simulation with double precision were used with different algorithms for the integration in time (4-order Runge-Kutta method with the fixed time step, RK4; 4-5-order Runge-Kutta method with adaptive time step, RK45; the Split-Step Fourier method with a midpoint finite-difference approximation for the nonlinear terms (following [30]), SSF, and its combination with the 4-order Runge-Kutta method, SSF&RK4). The full de-aliasing of the codes was provided. Besides, some simulations were verified by means of the ultra-high precision calculations using arbitrary-precision arithmetic (with 256-bit long mantissa, which is roughly equivalent to 80 decimal digits) and with the iterations in time performed by the 4-5-order Runge-Kutta method. We should mention that the problem of stable numerical simulation of the Akhmediev breathers was also tackled in [28] in a different aspect.

The first example shown in Fig. 1 corresponds to the case of the steepness $\varepsilon = 0.11$ but a longer perturbation, $N_x = 10$, which corresponds to $R_{BFI} \approx 3.11$; thus 3 unstable modes may develop. The evolution of selected Fourier amplitudes is shown. The initial condition is specified in the form of the exact breather solution (5); the targeted focusing time is $t = 0$. The adjacent to the dominant wave Fourier sidebands have the amplitude $1 \cdot 10^{-8}$ of the amplitude of the dominant harmonic (the $\pm 1$ sidebands, see the leftmost side of Fig. 1). For convenience, the amplitude of the spatial Fourier harmonics are normalized with respect to the amplitude of the dominant one; they will be referred to as $a_n$, where $n$ counts the sideband on the discrete wavenumber axis. Hence, initially $a_0 = 1$ and $a_{\pm 1} = 1 \cdot 10^{-8}$. Due to the computer roundoff at the level $1 \cdot 10^{-16}$, the other sidebands prescribed by the solution (6) are initially below the precision of the Fourier transform shown in the figure.

The evolution of the analytic solution (6) for $a_n$, $n = 0, \pm1, \pm2, \pm3$, is given in Fig. 1 by the plane lines, while the data of the numerical simulations is given by the curves with symbols, see the legend. The simulated waves follow the analytic solution for a few hundred wave periods. The modulations at the ±2 and ±3 sidebands appear from the null values at the appropriate times and also follow the analytic solution. The focusing of the analytic and numerical solutions occurs at the same time. The lower and higher frequency sidebands $a_{-n}$ and $a_{+n}$, $n = 1, 2, 3$, at each instant have similar amplitudes due to the symmetry of the NLS equation. However, about one hundred wave periods later the dynamics changes, as the sidebands ±2 gain most of the energy of the modulation and at $t \approx 350 T_0$ the most developed modulation is twice shorter than the domain size. Later on, several unstable modes accrue significant amounts of energy; the wave dynamics becomes complicated and strongly irregular, see $t > 1000\, T_0$.

The simulation presented in Fig. 1 may be improved, for example, by using a more accurate method for the integration in time. However, it is clear that due to the presence of several unstable Fourier harmonics, a weak perturbation of one of them may eventually cause the excitation of the other unstable modes. If the magnitude of the perturbation to the initial condition, $a_{\pm 1}$, is further reduced from $10^{-8}$ to $10^{-15}$, then the modulation which has the length half of the simulation domain, $L/2$, develops faster and completely changes the picture of evolution in Fig. 1. To remedy this effect, in the simulations below we consider the case of a shorter perturbation length, $N_w = 4$, which provides the single-mode regime of the modulational instability.

Fig. 2 presents the results of the simulation of the NLS equation with parameters $\varepsilon = 0.11$, $N_w = 4$, when the perturbation at the ±1 sidebands amount to $10^{-15}$ of the dominant harmonic amplitude. Despite such an incredibly small level of the deterministic perturbation (noise at other sidebands seem to exceed its magnitude, see the leftmost side in Fig. 2a), the simulated dynamics (the curves with symbols) follows the exact analytic solution (the plane lines) with high accuracy for more than a thousand wave periods. However after the first focusing at $t = 0$ the amplitudes $a_{\pm 1}$ do not continue decreasing below the value of about $5 \cdot 10^{-8}$. At some moment the dynamics reverses and instead of demodulation, a new cycle of the wave self-modulation starts. This process continues repeatedly at longer times.

The respective Fourier phase of the modulation to the dominant wave (i.e., the value of $\theta$ in (6)) is important. For the chosen parameters $\varepsilon = 0.11$ and $N_x = 4$ the phase $\theta_{-\infty} \approx -0.20\pi$. It is convenient to consider the dynamic phase, which may be defined for the adjacent to the carrier wave sidebands as

$$\Theta_1 = \arg\left(\hat{q}_1 \hat{q}_{-1} \hat{q}_0^* \hat{q}_0^*\right). \tag{12}$$

For the simulation shown in Fig. 2a the function $\Theta_1$ exhibits fast evolution from the value $2\theta$ to $-2\theta$ at the moments of focusing $t/T_0 = 0$, $t/T_0 \approx 900$, ..., and abrupt jumps when the defocusing changes to focusing (Fig. 2b). The oscillations at the initial stage of the evolution give the evidence that initially the dynamics is not coherent, the favorable for the modulational growth phases are settled naturally.

The other nearest sidebands ±2, ±3 follow the dynamics of the first sidebands ±1 (Fig. 2a). They start to grow from the noise level, increase and then return to the noise; during the first focusing event they coincide with the analytic solution both in amplitude and phase. The overall picture of the evolution in the wavenumber Fourier space does not exhibit noticeable accumulation of energy at any other scales (Fig. 2c).

The use of the other numerical algorithms with double precision and better resolution in space and time do not help to improve significantly the recurrence shown in Fig. 2 (in particular, to increase noticeably the period of the partial recurrence). The dynamics of $\Theta_1$ when the demodulation turns to a new self-modulation (Fig. 2b) may be qualitatively different if another algorithm for the iterations in time is used. A better recurrence is obtained the ultra high precision simulations, when the initial amplitude of $a_{\pm 1}$ was taken as small as $1 \cdot 10^{-23}$. The period between two consecutive focuses increases about 1.5 times, and the sideband amplitudes in the demodulation stage drop to smaller values about $5 \cdot 10^{-12}$. However the repeating cycles of the modulation and demodulation phenomena are also observed.

The time reversibility was also checked by virtue of the backward simulation starting from the numerically calculated waves at the moment $t = 0$ (Fig. 3). The backward simulation coincides with the forward wave dynamics for some time; it exhibits wave demodulation until the sidebands become too small, $a_{\pm 1} \approx 5 \cdot 10^{-8}$ (Fig. 3a). Then the backward demodulation process changes to the new focusing cycle similar to the simulation in the positive direction of time. Thus, the picture of the backward simulation $t < 0$ in Fig. 3a may be obtained by the mirror reflection of the curves for $t > 0$ with respect to the time reference, $|a_j(-t)| \approx |a_j(t)|$. The same symmetry holds for the phase (Fig. 3b), though the sign of $\theta$ changes, i.e., $\theta(-t) \approx \theta(t)$.

To conclude, the disagreement between the analytic breather solution of the NLS equation and its numerical simulation presented here (Fig. 2) can occur due to the following reasons (see also [28]):

i) due to the different dynamical features of the continuous partial differential equation and its discrete counterpart;

ii) due to the inner inaccuracy of the employed algorithms which solve the equations;

iii) due to the computer roundoff.

Though we do not identify the particular mechanism which limits the accuracy of the numerical calculation of the NLS breather, we clearly demonstrate the following properties:

1) the repetition of the modulation-demodulation phenomena is not robust with respect to the algorithm of the calculation. The quasi-recurrence period grows and thus the recurrence improves if the computation accuracy increases;

2) the numerical simulation of the modulation-demodulation cycle is isotropic in time. The dynamics is not time-reversible if the sidebands in the Fourier space reach some small level of magnitudes.

Therefore we may conclude with the following important statement, that the effect of incomplete recurrence of the NLS breather which is observed in the performed numerical simulations is mainly related to the *computer roundoff error*. The limiting effects of the first two listed above possibilities, i) or ii), is of a lesser importance.

## 5. Simulations of the fully nonlinear potential Euler equations

In this section the numerical simulations of the weakly perturbed Stokes waves within the frameworks of the primitive hydrodynamic equations are presented. The set of equations for potential surface gravity waves in infinitively deep uncompressed inviscid ideal fluid may be represented in the form [31]

$$\frac{\partial \eta}{\partial t} = -\frac{\partial \Phi}{\partial x}\frac{\partial \eta}{\partial x} + \left(1 + \left(\frac{\partial \eta}{\partial x}\right)^2\right)\frac{\partial \varphi}{\partial z}, \quad \text{at} \quad z = \eta, \tag{13}$$

$$\frac{\partial \Phi}{\partial t} = -g\eta - \frac{1}{2}\left(\frac{\partial \Phi}{\partial x}\right)^2 + \frac{1}{2}\left(\frac{\partial \varphi}{\partial z}\right)^2\left[1 + \left(\frac{\partial \eta}{\partial x}\right)^2\right], \quad \text{at} \quad z = \eta, \tag{14}$$

$$\frac{\partial^2 \varphi}{\partial x^2} + \frac{\partial^2 \varphi}{\partial z^2} = 0, \quad z \leq \eta, \tag{15}$$

$$\frac{\partial \varphi}{\partial z} \to 0, \qquad z \to -\infty. \qquad (16)$$

Here $Oz$ is the upward vertical axis, $\eta(x, t)$ is the surface displacement, $\varphi(x, z, t)$ is the velocity potential, and $\Phi = \varphi(x, z = \eta, t)$ is the surface velocity potential. The equations (13)-(16) are solved by a pseudo-spectral code without any approximation when formulated in conformal variables [32]. This approach is used in this study as the main framework. An ultrahigh-precision version of the code was implemented and simulated in addition using double-double arithmetic providing 128 bits of precision (about 30 decimal digits). The High-Order Spectral Method (HOSM) [33] which takes into account up to 7-wave interactions is also used for the simulations with the purpose to estimate the robustness of the simulated wave dynamics.

The initial condition is specified in the form of numerically exact Stokes waves with the crest-to-trough steepness $\varepsilon = 0.11$, $\varepsilon = k_0 H/2$, $H$ is the wave height; four wave periods are simulated in a periodic domain, $N_x = 4$. The dominant wave parameters are the same as in the simulation of the NLS equation in Sec. 4, Fig. 2. Similar to the simulations discussed above, the modulation is introduced with the help of the spectral sidebands $a_{\pm 1}$ of a very small magnitude $10^{-15}$, comparable with the level of numerical noise. The simulation is represented in Fig. 4, which is similar to the Fig. 2, but now equations (13)-(16) are solved. The modulational growth observed in the simulation of the Euler equations is noticeable slower than predicted by the NLS theory (the lines with symbols versus the plane lines in Fig. 4a); the first focusing occurs with the delay of about 650 wave periods. The dynamic phase behaves qualitatively similar to the NLS solution, but the quantitative difference is obvious (Fig. 4b). It may be seen that the second sidebands ±2 grow faster than the first ones (±1) similar to the NLS framework (Fig. 2a). The third high frequency sideband +3 corresponds to the same wavenumber as the first low frequency sideband of the second harmonic, $k_0 + 3\Delta k = 2k_0 - \Delta k$, (in the discussed case $\Delta k = k_0/4$). The latter dominates in magnitude, hence the amplitude $a_{+3}$ cannot be evaluated from the spatial Fourier transform. The sideband −3 corresponds to the same wavenumber as the first sideband of the zeroth harmonic, $k_0 - 3\Delta k = k_0 - k_0 + \Delta k$. Therefore the amplitude $a_{-3}$ cannot be evaluated from the wavenumber Fourier transform either. That is why the evolution of only two sidebands are shown in Fig. 4. The differences between the low and high frequency sidebands, $a_{-2}$ and $a_{+2}$ respectively, are well seen in Fig. 4a; this feature is absent in the NLS framework.

Similar to the weakly nonlinear theory, the initial perturbation causes the development of modulation. The sidebands withdraw energy from the dominant wave, and attain the saturation at about $650T_0$; then the process seems to reverse, but the recurrence is obviously not complete. After some period of demodulation, the instability commences again resulting in a new cycle of the modulation-demodulation, and a quasi periodic dynamics is observed (a longer simulation is shown in Fig. 5) similar to the case of the NLS equation shown in Fig. 2a. The difference between the evolutions of the amplitudes $a_{+2}$ and $a_{-2}$ is striking when the modulation is suppressed after the focusing: $a_{+2}$ remains almost constant for some time while $a_{-2}$ drops to the values which are at least one order of magnitude smaller, and experiences more irregular fluctuations during this phase.

It is crucial, that the amplitudes of the sidebands during the stage of demodulation are significantly larger than in the case of the NLS equation (cf. Fig. 4a and Fig. 2a). Moreover, the dynamics of the Fourier amplitudes shown in Fig. 4 is stable with respect to the use of other accurate algorithms for the simulation. In particular, the results of three different long-term simulations are shown by different colours in Fig. 5: the conformal Euler equations solved by the RK4 and RK45 algorithms, and the HOSM with RK4 simulation. The evolution of selected Fourier amplitudes and the dynamics of $\Theta_1$ are given in Fig. 5a and Fig. 5b respectively. The simulation of the conformal Euler equations with the ultrahigh precision, when the initial perturbation is characterized by even smaller amplitude $a_{\pm 1} = 1 \cdot 10^{-23}$, supports this result as well. All the simulations report the same quasi-recurrent dynamics with the same period of the modulation 'breathing'. Some small lead of the evolution simulated by the HOSM may be found compared with the simulations of the Euler equations in conformal variables. A small deviation is accumulated by the simulation with the ultrahigh precision (circles in Fig. 5). The dynamic phase undergoes smooth evolution and unambiguously determines the modulation or demodulation stage. We remark that this property could be used for the purpose of forecasting of rogue waves as it is selective with respect to the approaching or diminishing extreme event for the same given shape of the envelope.

Fig. 4c illustrates how the presence of the bound Stokes wave components complicates the evolution of the spatial Fourier transform of the modulated surface displacement. Compared to Fig. 2c, where the envelope function is analyzed, the pattern in Fig. 4c represents the evolution of a comb-shaped profile. When the waves focus, the modulation becomes so strong that the peaks of the multiple harmonics can be hardly distinguished in the momentary Fourier transform.

The examination of the time-reversal property has been performed similar to the case of the NLS equation. The backward simulation started from the moment after the first 'breath' of the modulation $t \approx 1000 T_0$, see Fig. 6. For some time of the backward simulation the wave train experiences demodulation exhibiting the full reversibility of the evolution, though later the demodulation turns to a new self-modulation cycle. Essentially, the first switch of the demodulation/modulation regime for the backward simulation takes place when the magnitudes of the sidebands are noticeably smaller than for the forward simulation (Fig. 6a). For example, the local minimum of $a_{\pm 1}$ is about $5 \cdot 10^{-3}$ at $t \approx 1000 T_0$ and about $2 \cdot 10^{-5}$ at $t \approx 170 T_0$ for the backward simulation. The difference for the second sidebands is even more pronounced in Fig. 6a. The evolution of the dynamic phase in the forward and backward direction, calculated by means of the same numerical algorithm, is qualitatively different (Fig. 6b). Thus the observed envelope dynamics is not symmetric with respect to the instant of the maximal focusing, $t \approx 653 T_0$, in contrast to the case of the NLS equation shown in Fig. 3. This observation helps to conclude that the incomplete demodulation after the wave train focusing, which is found in the fully nonlinear simulations of the potential Euler equation, is not a numerical artifact due to the insufficient accuracy of the simulation. The physical interpretation of the observed effect is discussed in the next section.

## 6. Discussion

The physical mechanisms responsible for the observed non-recurrent dynamics in the simulation of the Euler equations displayed in Fig. 4 is examined in this section with the help of the windowed space-time Fourier transform of the surface displacement $\hat{F}\{\eta\}$. The Hanning window function of time $M(t) = 0.5 - 0.5 \cos(2\pi t/W)$ of the length $W$ is applied to select the time intervals. This function swamps the data at its edges and hence improves the efficiency of the Fourier transform along the time axis (as the original data is not periodic in time). The space-time Fourier transform amplitudes $S$ are determined as

$$S(k,\omega) = 2\left|\hat{F}\{\eta(x,t)M(t-\tau)\}\right|, \quad \omega > 0. \tag{17}$$

The Fourier amplitudes for the wavenumbers, $S_k$, are then calculated after the integration by frequencies,

$$S_k^2 = \int_0^\infty S^2(k,\omega)d\omega. \tag{18}$$

The frequency Fourier transforms for the waves with positive and negative phase velocities, $S_{\omega+}$ and $S_{\omega-}$, are obtained by the integration along the positive and negative wavenumbers respectively,

$$S_{\omega+}^2 = \int_0^\infty S^2(k,\omega)dk, \qquad S_{\omega-}^2 = \int_{-\infty}^0 S^2(k,\omega)dk. \tag{19}$$

Hereafter, waves with $k > 0$ will be referred to the following waves, while waves with $k < 0$ – to the opposite waves.

The evolution in time of the maximum of the surface displacement, which corresponds to the simulation in Fig. 4, is shown in Fig. 7 with the black curve. The selections performed by the window function $M$ are shown with the filled red areas. The selections correspond to the first focus, $t = 653T_0$, and also to two equidistant from this moment instants $t = 322T_0$ and $t = 1000\ T_0$. The latter corresponds to the local minimum of the Fourier component $a_{\pm 1}$, see Fig. 4a. The moment $t = 1\,200\ T_0$ corresponds to the stage of a resuming modulational growth. The frequency Fourier transforms which correspond to these four instants are shown in Fig. 8a,b for the opposite and following waves respectively. These spectra are too difficult for interpretation, therefore the corresponding space-time Fourier transforms are plotted in Fig. 9. In Fig. 9 the functions $S_k$, $S_{\omega-}$ and $S_{\omega+}$ are also plotted, respectively, at the up, left and right sides of each panel. The snapshots of the surface displacement are given in the upper-left quarters by magenta curves. The deep-water linear dispersive curve $\omega^2 = g|k|$ is plotted in Fig. 9 by the dotted cyan lines for the reference.

The uniform Stokes wave is characterized by the dominant wavenumber $k_0$ and the peak frequency $\omega_p$, which slightly exceeds the linear frequency $\omega_0 = (gk_0)^{1/2}$ due to the effect of nonlinearity. The nonlinear bound (phase-locked) waves are represented in the $(k, \omega)$ Fourier space by the peaks located at $(nk_0, n\omega_p)$, $n = 0, 2, 3, \ldots$. Besides, the modulation yields the series of sidebands which adjoin to the dominant peak and to the bound wave components. This brief description helps to interpret Fig. 9a, when the modulation is still small. Due to the periodic boundary conditions the wavenumber Fourier transforms are discrete functions (see e.g. $S_k$ at the upper side of Fig. 9a). The tails of the Fourier transform $S(k, \omega)$ decay slower along the frequency axis (see also $S_{\omega+}$ at the right hand side of Fig. 9a). This dependence is determined by the mask function $M(t)$, it limits the minimal Fourier amplitudes which can be resolved by the processing (see the green filled area in Fig. 8b, which corresponds to $S_{\omega+}$ for the first case $t = 347T_0$). It is important to note that the carrier wave peaks and their sidebands form straight parallel lines in the $(k, \omega)$ plane, what reveals the occurrence of a strongly coherent group (see also in [29]). The inclination of these lines specifies the velocity of the coherent group. The function $S_{\omega+}$ for $t = 347T_0$ (the green filled area in Fig. 8b) exhibits the Stokes wave components and the small modulation components. The values of function $S_{\omega-}$, which characterizes the opposite waves, are below the limits of Fig. 8a at this instant.

The blue curve in Fig. 8a,b and the plot in Fig. 9b correspond to the stage of the first focus of the modulated waves (the second selection in Fig. 7). The plot of $S_{\omega+}$ in Fig. 8b contains many frequencies which build up a quite complicated spectrum. In Fig. 8c the zoomed interval of $S_{\omega+}$ around the dominant peak is given. However, Fig. 9b is more useful for the interpretation. The lobes which correspond to the different bound wave components still may be easily distinguished in Fig. 9b, though the developed modulation results in the appearance of the long tails of sidebands adjacent to each of the Stokes wave harmonics. The modulation is so strong that the lobes which correspond to different bound components overlap in Fig. 8b, and the Fourier amplitudes for the negative wavenumbers $S_{\omega-}$ obtain non-zero values (better seen in Fig. 8a). All together they form the shape of the envelope of the focused train in space and time. It is remarkable that the energy spots in Fig. 9b, which belong to different bound wave components, are still parallel, thus the simulation of the Euler equation exhibits a coherent wave structure similar to the weakly nonlinear NLS breather despite the strong modulation (see the magenta curve in Fig. 9b) and large steepness $k_0 \max \eta \approx 0.22$.

The instant $t = 1000\,T_0$ corresponds to the stage of a minimal modulation between the two focuses (Fig. 7, Fig. 9c). The moment of the first focus, $t = 653\,T_0$, is exactly in the middle between the instants $t = 347\,T_0$ and $t = 1000\,T_0$. If the wave modulation is a fully recurrent process, Fig. 9c would be a duplicate of Fig. 9a. However, the differences between these figures are quite obvious. The most evident is the appearance of new waves which belong to the dispersive curve and propagate in the opposite ($k < 0$) and following ($k > 0$) directions. They are generated due to the group-wave resonance by virtue of the strongly modulated dominant and the zeroth harmonics respectively, see [29]. The Fourier mode which corresponds to the opposite wave with the wavenumber $-k_0/2$ is represented in Fig. 8a by a peak at $\omega \approx \omega_0/\sqrt{2}$; it remains about the same for the instants $t/T_0 = 653, 1000, 1200$ (better seen in the inset in Fig. 8a). The corresponding spots of energy are also well seen in Fig. 9b–d. The Fourier amplitude may be estimated in terms of the steepness as $O(\varepsilon^{-6})$ (see the inset in Fig. 8a). The following wave which is generated by the breathing group has the wavenumber close to $4k_0$, it is almost invisible in Fig. 9c. Indeed, in the limit of vanishing nonlinearity the zeroth harmonic of the wave group with the velocity $\omega_0/2/k_0$ should cross the linear dispersion curve at ($4k_0, 2\omega_0$). The corresponding peak can be observed directly neither in $S_k$ nor in $S_{\omega+}$ due to the overlap with other more energetic Fourier components. A selecting mask in the ($k$, $\omega$) plane may be applied to calculate the magnitude of this wave component, it is of the order $O(\varepsilon^{-7})$ in the simulated case. Besides, the harmonics which correspond to the combinations of the dominant wave and the generated free waves may be also found in Fig. 9c,d.

A thorough comparison of the curves $S_{\omega+}$ in Fig. 8c for the instants $t = 347\,T_0$ (the green filled area) and $t = 1\,000\,T_0$ (thin black) leads to the conclusion, that before the focusing the second ($\pm 2$) sidebands have frequencies about $0.75\omega_0$ and $1.27\omega_0$, which are distinctly different from the prescribed by the dispersive law; thus these sidebands are phase-locked modes of the modulation. The frequencies allowed by the set of the discrete wavenumbers in accordance with the modified dispersion relation (20) are marked with the vertical dashed lines in Fig. 8; the frequency discretization in Fig. 8 is about $0.006\omega_0$. In contrast, after the first focusing event most of the energy of the second high frequency sideband (i.e., of $a_{+2}$, which correspond to the wavenumber $3k_0/2$) is placed at the frequency for free waves (the thin black peak coincides with the vertical dashed line at $1.24\omega_0$; note that the peak is shifted leftward with respect to the green peak), its amplitude is estimated as $O(\varepsilon^{-4})$. In turn, the energy in the low frequency part of $a_{-2}$ with the wavenumber $k_0/2$ is distributed between a free wave with the frequency $0.71\omega_0$ and a phase-locked modulation component with the frequency about $0.77\omega_0$ which does not satisfy the dispersion relation (note that the bound component is shifted rightward compared to the location before the focusing, shown with the green line in Fig. 8c; the peak splitting is also seen in Fig. 9c). The corresponding modification of the frequency Fourier transform is also observed at the zeroth and the second harmonics in Fig. 8b. Meanwhile the dominant peak and the first-order sidebands look rather similar at the instants before and after the focusing event (Fig. 8c). The described evolution seems to be similar to the partial recurrence observed in the numerical simulations of the high-order NLS equation in [25].

The free and phase-locked wave components which correspond to the same allowed discrete wavenumber become distinguishable due to the decomposition in the frequency Fourier domain and may be interpreted with the help of the space-time Fourier plots in Fig. 9.

When the second cycle of the self-modulation starts (the instant $t = 1200\,T_0$), the second sidebands are represented in the frequency domain in Fig. 8c by multi-humped shapes of $S_{\omega+}$, which include the contributors from the coherent and incoherent parts, see also Fig. 9d. When the modulation develops further, the coherent sidebands dominate. The resolution of the frequency Fourier transform does not allow discerning the difference between the free and coherent components which correspond to the first sidebands ±1 (i.e., with the wavenumbers $k_0 \pm k_0/4$).

We should add that for the proper description of the dispersive properties of the small-amplitude waves propagating over an intense dominant wave, a modified dispersion relation was used in Fig. 8, according to the weakly nonlinear theory for coupled waves (e.g., [34]). Then the dispersive relation contains a nonlinear correction as follows,

$$\omega_{nl} = \sqrt{gk} + \varepsilon^2 \omega_0 \left(\frac{k}{k_0}\right)^p, \quad p = 1 \text{ if } |k| > |k_0|, \text{ and } p = 2 \text{ if } |k| < |k_0|, \quad (20)$$

where $\varepsilon = 0.11$ is the steepness of the initial condition. This correction is essential for the interpretation of the second high frequency sideband of the dominant wave (at $k = 3k_0/2$) in Fig. 8c. The excursion of the emerged free waves from the linear dispersion relation may be also noticed in Fig. 9b–d.

The described above excitation of free waves with wavenumbers the $k_0/2$ and $3k_0/2$ explains the evolution of the sideband amplitudes $a_{\pm 2}$ in the fully nonlinear simulations in Fig. 3a, 4a. The amplitude of the corresponding sideband does not decrease below the amplitude of the generated free wave. The amplitude of the high-frequency free wave is about one order larger than the amplitude of the low-frequency one.

The overall evolution of the frequency transform $S_{\omega+}$ with time is shown in Fig. 10 (a certain frequency interval is displayed). Note the crucial difference between Fig. 4c and Fig. 10, as in Fig. 4c the momentary wavenumber Fourier transform is shown which cannot distinguish phase-locked and free waves, while in Fig. 10 they become resolved due to the different frequencies. The different patterns before the first focusing event and between the two consecutive focuses illustrated in Fig. 10 are quite obvious, they agree with the description of Fig. 8. New frequencies obtain energy in the course of the strong self-modulation. These peculiarities determine the anisotropy of the modulation evolution in time (Fig. 5) and lead to the incomplete recurrence of the modulational instability.

## 7. Conclusion

In [15] the breather solutions of the NLS equation with the attribute of "appearance from nowhere and disappearance without a trace" were considered as perfect prototypes of the rogue waves. With the help of the fully nonlinear numerical simulations we have obtained the evidence that the breather solution of the potential hydrodynamic equations does not exist in the sense that it does not satisfy the criterion formulated in [15]. Namely, if a negligible modulation of a wave train starts to grow and at some moment causes the rogue wave event, the subsequent modulation disappearance is incomplete. In the course of the strong self-modulation of the Stokes wave other free waves emerge. The opposite and the following waves are generated due to the group-wave resonances of different orders of the nonlinearity as described in [29]. Besides, the effect of the generation of incoherent waves with the lengths comparable with the length of the dominant Stokes wave is found to be even more significant. The incomplete recurrence leads to the quasi-periodic breathing of the modulated wave train. In contrast to the simulations of the integrable NLS equation, where the departure from the exact breather solution was due to the hardly controllable effects of noise, the quasi-periodic dynamics of the fully nonlinear waves is robust. These mechanisms of generation of the waves with new lengths generalize the already known effects of wave emission in nonlinear optics related with the Cherenkov radiation and wave-soliton interaction [35-39].

In the study we have focused on the most significant situation when the waves are essentially steep to possess relatively fast nonlinear evolution, at the same time they do not reach the breaking limit, thus they attain the maximum amplification due to the self-modulation. The length of the perturbation is imposed to be four times larger than the wave length with the purpose to consider the simplest single-mode regime of the modulational instability. A few other cases with different wave parameters which support the single-mode regime of the modulational instability were also simulated ($\varepsilon = 0.07$, $N_w = 6$; $\varepsilon = 0.095$, $N_w = 5$; $\varepsilon = 0.146$, $N_w = 3$), two of them were described in less detail in [29]. The generation of new free waves and the quasi-periodic breathing dynamics of the modulation were observed in all those simulations.

On the other hand, the maximum amplitude of the incoherent waves generated by one cycle of the modulational instability is found to be four orders of magnitude smaller than of the dominant wave. Therefore in the majority of practically significant cases the water wave breathers which appear from nowhere may be considered as disappearing almost without a trace.

**Acknowledgements**
The support from RSF grant No. 16-17-00041 is acknowledged for the simulations with the standard double precision. The implementation of the solvers which employ the custom high-precision arithmetic and the related simulations were supported by RSF grant No. 14-17-00667. AS is grateful to E.N. Pelinovsky and V.I. Shrira for stimulative discussions and to A. Armaroli for providing important bibliographic references.

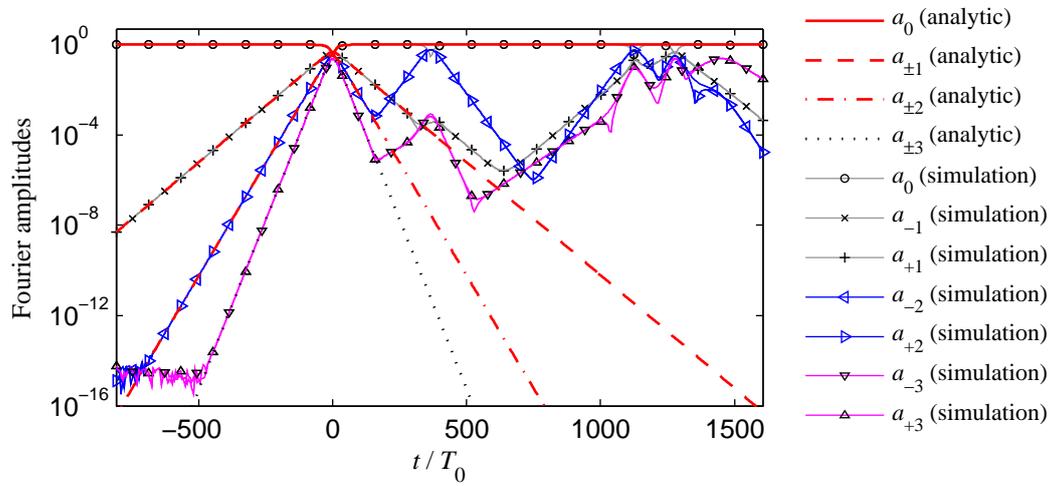

Fig. 1. The evolution of selected Fourier amplitudes in the numerical simulations of the NLS equation for the parameters $\varepsilon = 0.11$, $N_x = 10$ (RK4, the spatial resolution is $2^9$ points, the time step $dt = 2^{-7}$). The result of numerical simulation is compared with the exact analytic breather solution (plane lines).

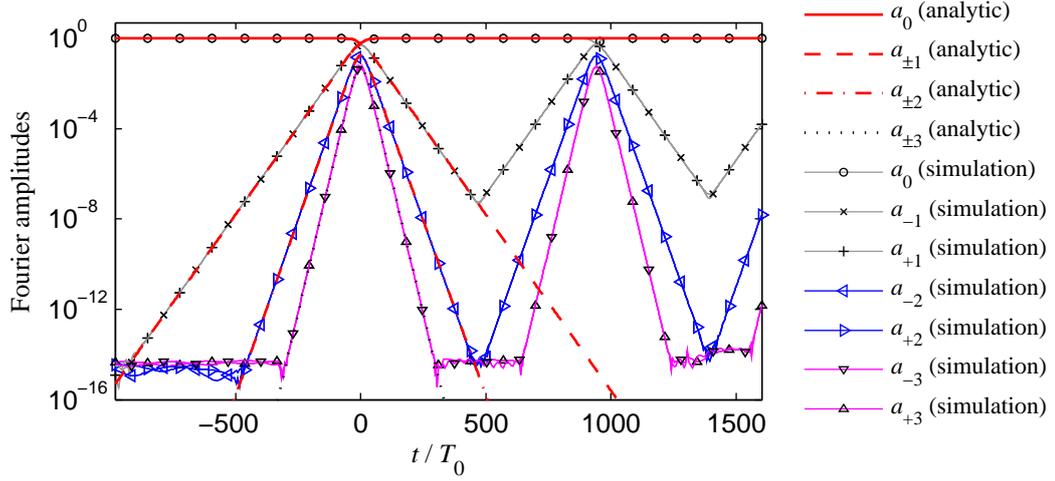

(a)

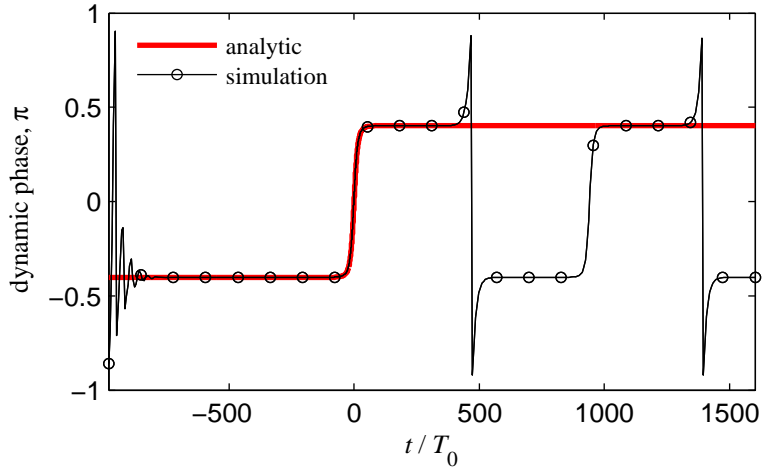

(b)

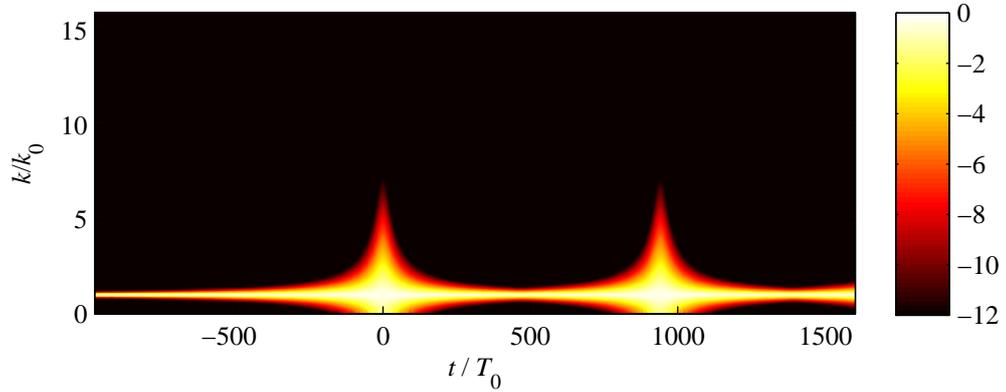

(c)

Fig. 2. The evolution of selected Fourier amplitudes (a), of the dynamic phase for the first sidebands, $\Theta_1$ (b), and of the overall spatial Fourier transform (c) in the numerical simulation of the NLS equation for the parameters $\varepsilon = 0.11$, $N_x = 4$ (RK4, $2^8$ points, time step $dt = 2^{-9}$). The result of numerical simulation is compared with the exact analytic breather solution. The decimal logarithm of the Fourier amplitudes is shown in panel (c) with the color for the range from 1 to $1 \cdot 10^{-12}$.

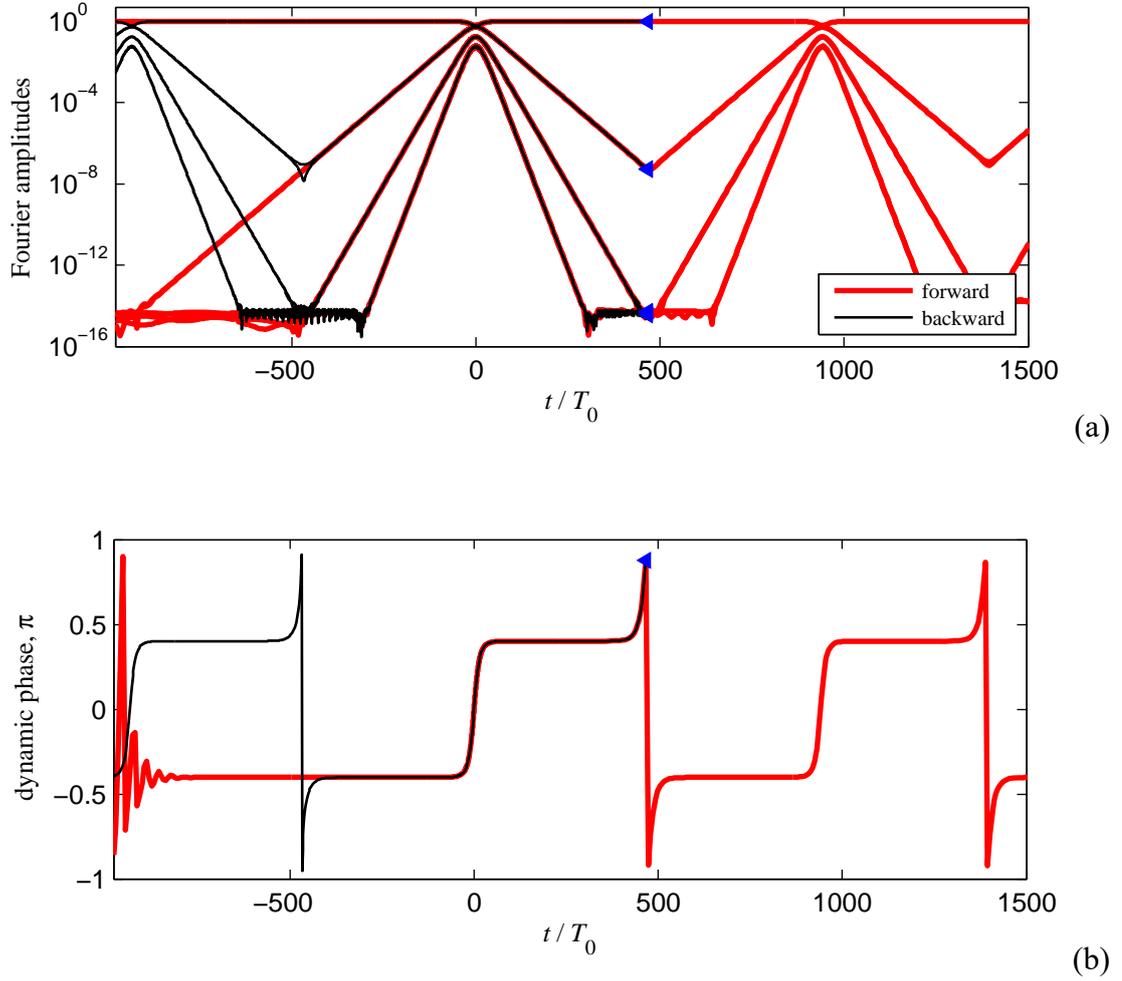

Fig. 3. The backward simulation of the NLS equation (the thin black curves) compared with the forward simulation (the thick red curves, same as in Fig. 2): the evolution of selected Fourier amplitudes (a) and of the dynamic phase $\Theta_1$ (b). The starting point of the backward simulation is marked with triangles.

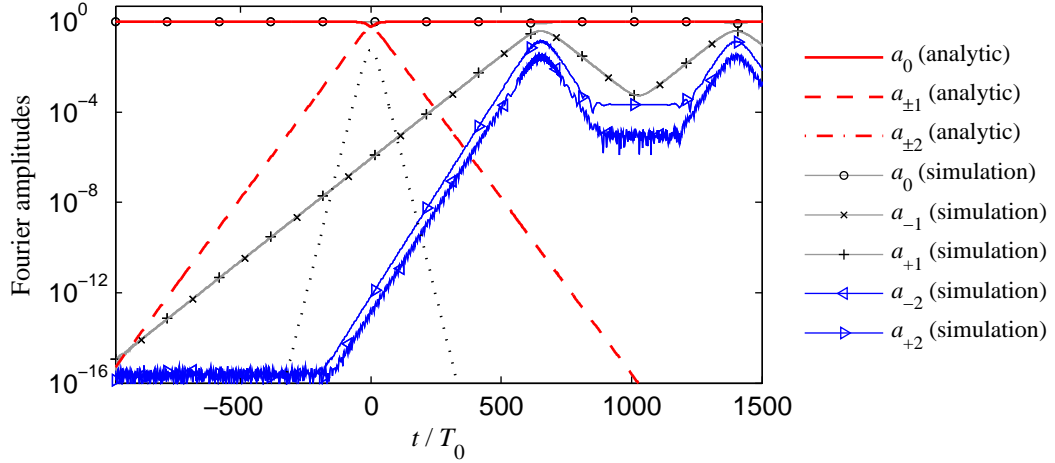

(a)

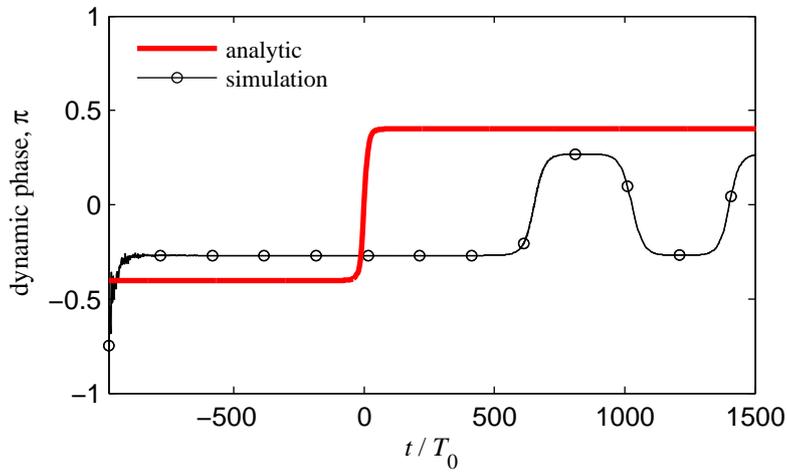

(b)

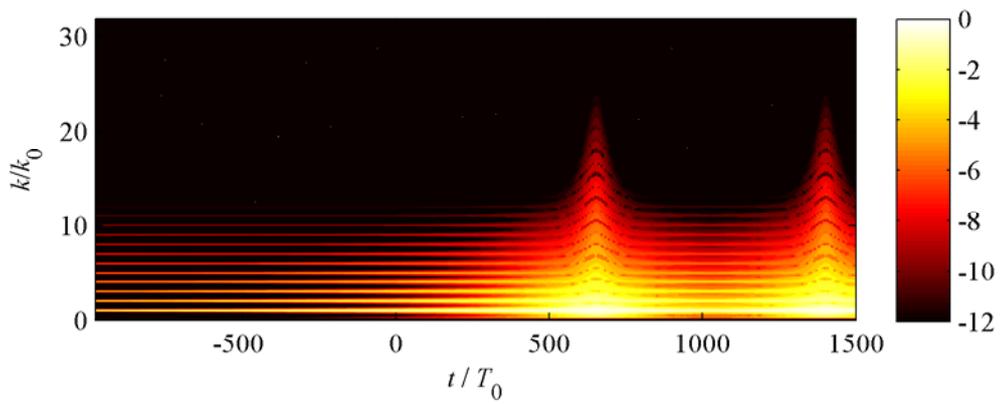

(c)

Fig. 4. Similar to Fig. 2, but the fully nonlinear simulation of the Euler equations is displayed (RK4, $2^9$ points, time step $dt = 0.04$). The result of numerical simulation is compared in (a, b) with the exact analytic NLS breather solution (plane lines).

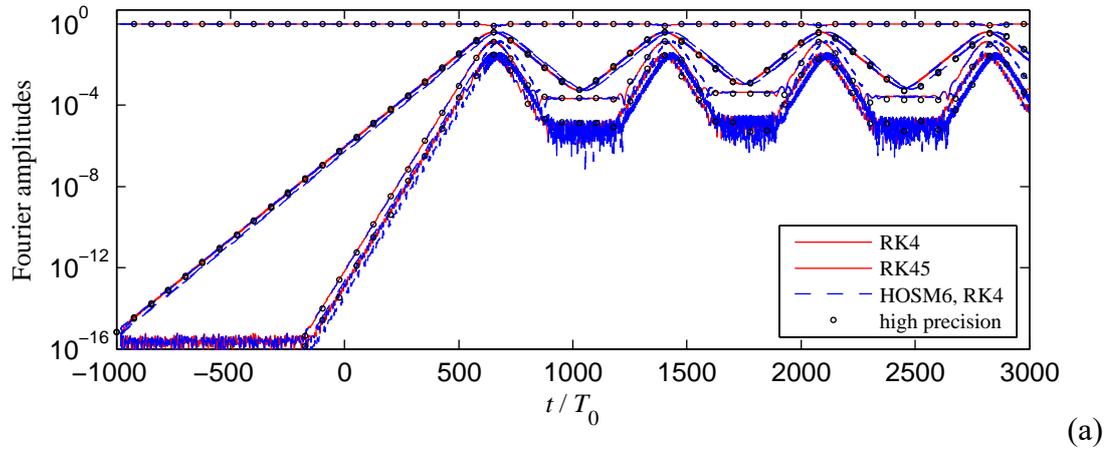

(a)

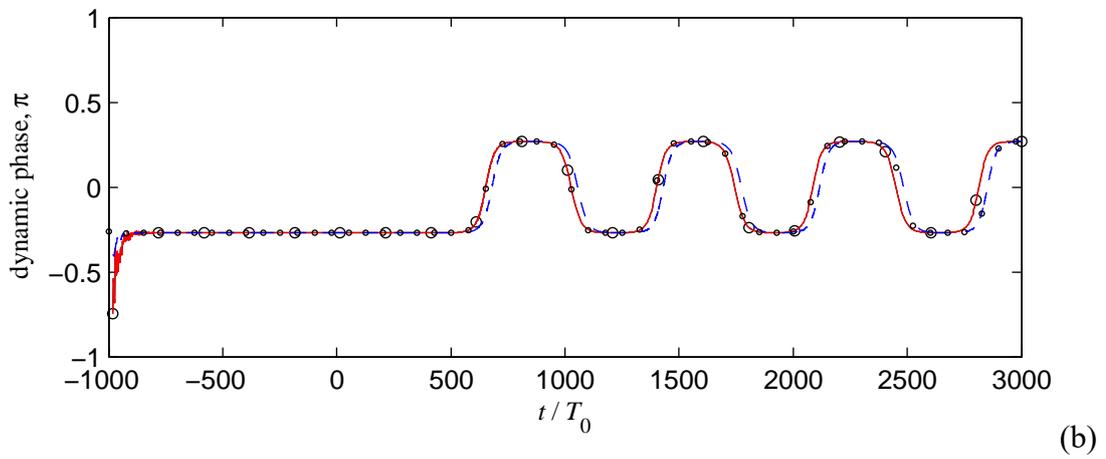

(b)

Fig. 5. Same as in Fig. 4a,b, but longer simulations are performed by means of three codes: the Euler equations in conformal variables iterated with the help of the RK4 and RK45 methods, the HOSM simulation by the RK4 method, and the simulation of the conformal Euler equations with the ultra-high precision (see the legend).

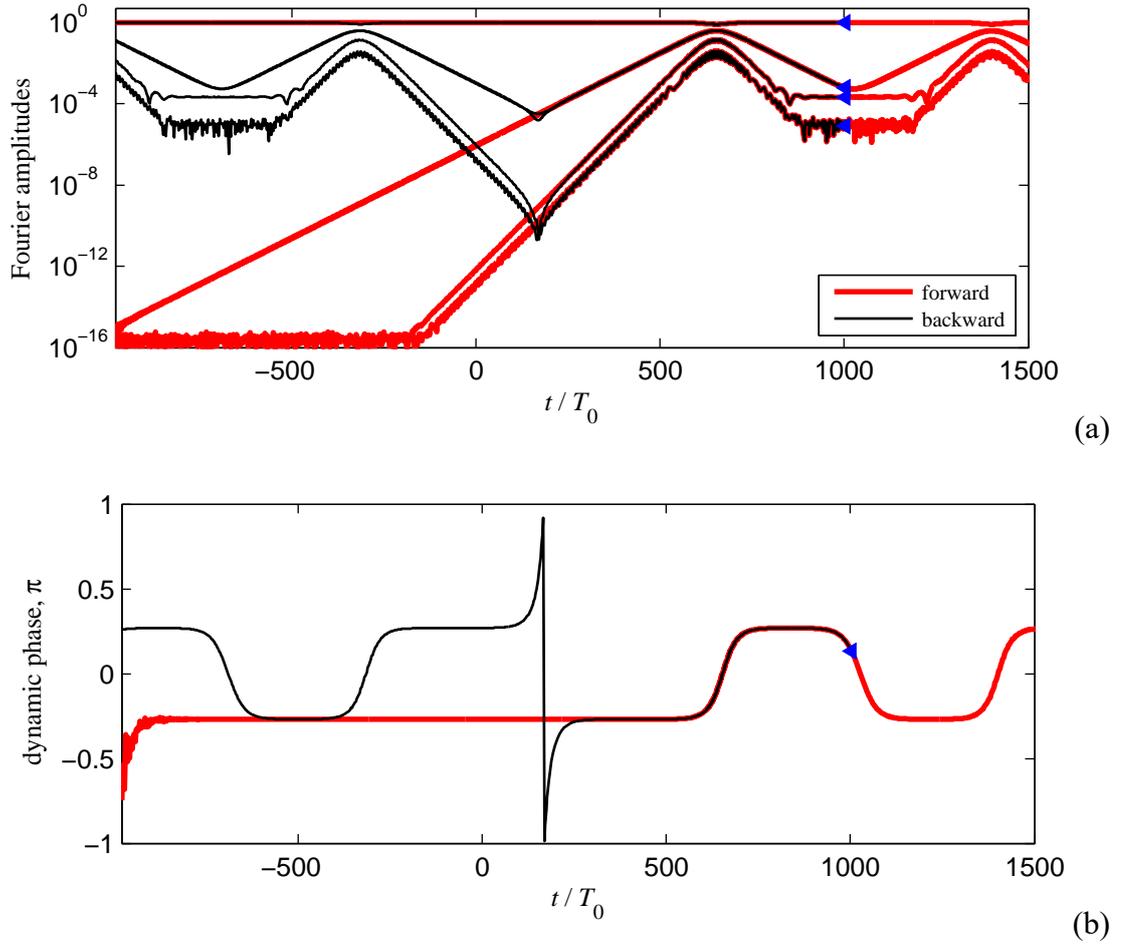

Fig. 6. The backward fully nonlinear simulation (the thin black curve) compared with the forward simulation (the thick red curve, same as in Fig. 4): the evolution of selected Fourier amplitudes (a) and of the dynamic phase $\Theta_1$ (b). The starting point of the backward simulation is marked with triangles.

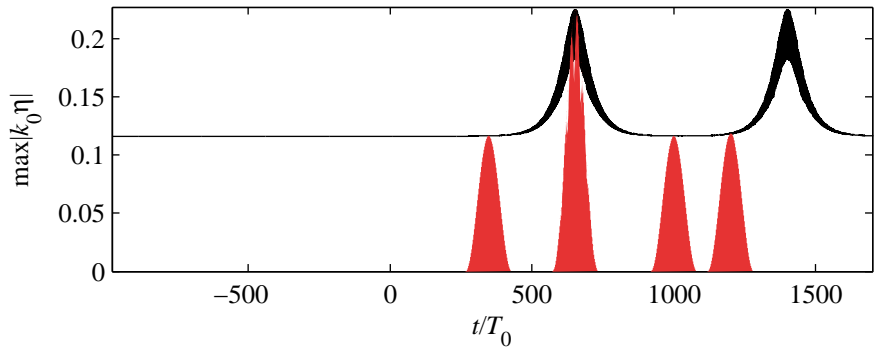

Fig. 7. The evolution of the maximum of the surface displacement and the window mask $M$ at the instants $t/T_0 = 322, 653, 1200, 1400$ for the fully nonlinear simulation shown in Fig. 4.

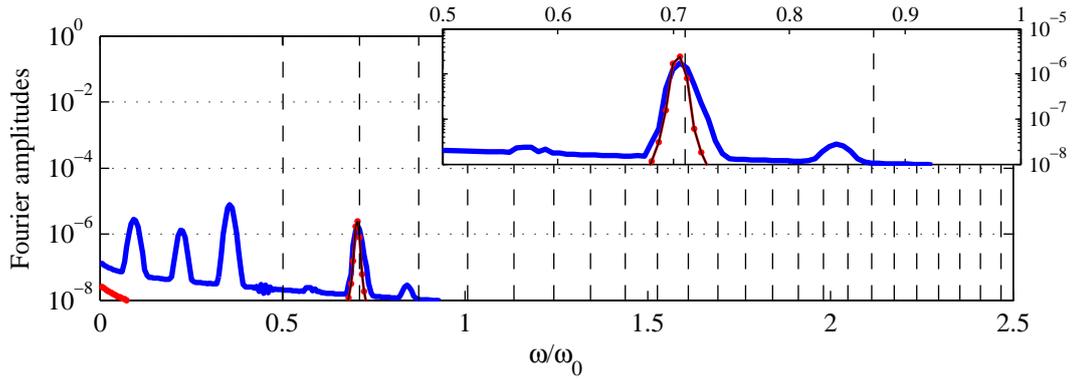

(a)

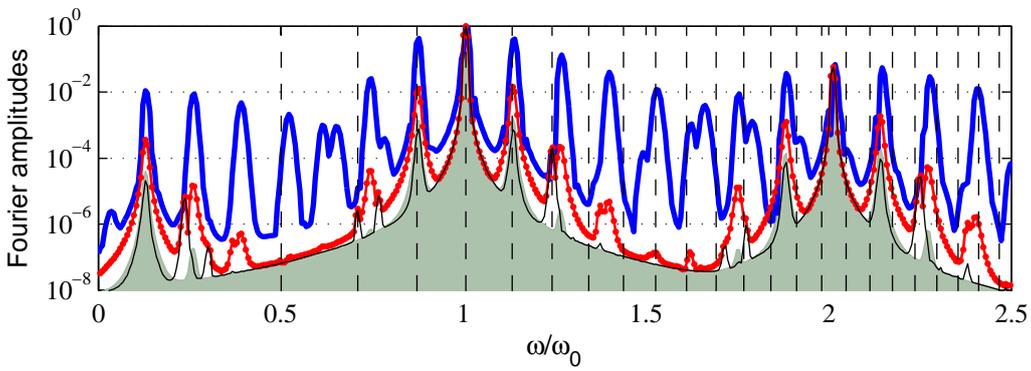

(b)

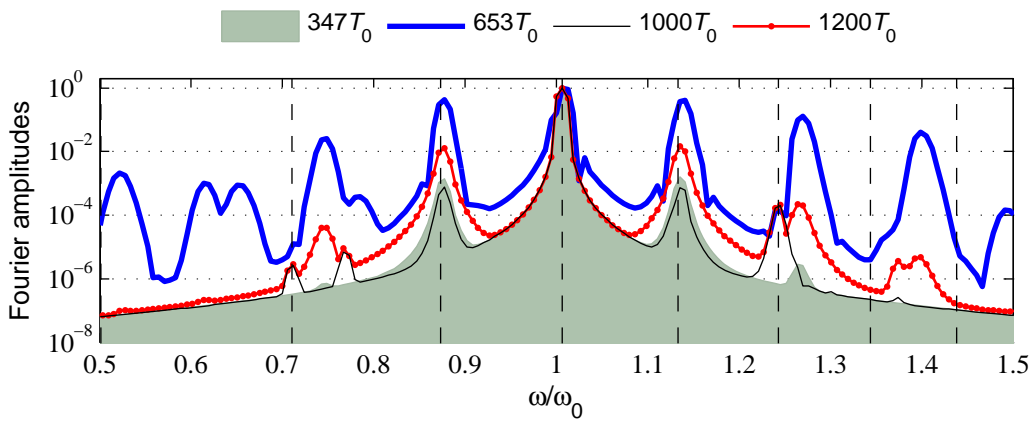

(c)

Fig. 8. The frequency Fourier transform amplitudes which correspond to the samples shown in Fig. 7: for the opposite waves, $S_{\omega-}$ (a) and the following waves, $S_{\omega+}$ (b). In panel (c) a part of $S_{\omega+}$ is given in the expanded scale. The vertical lines show the allowed frequencies $\omega_{nl}$ of weakly nonlinear free waves.

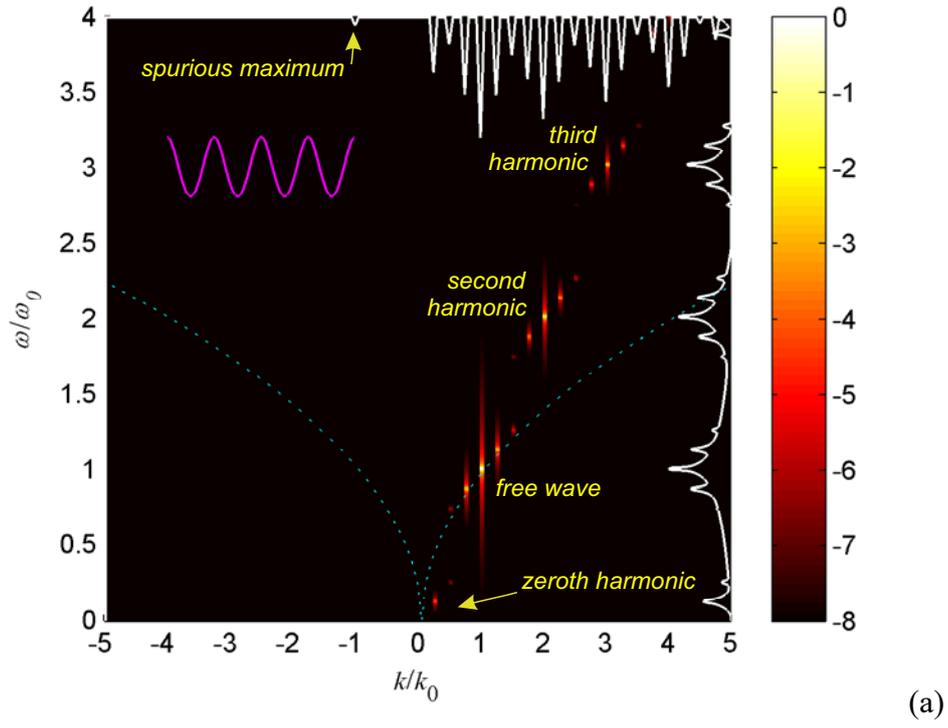

(a)

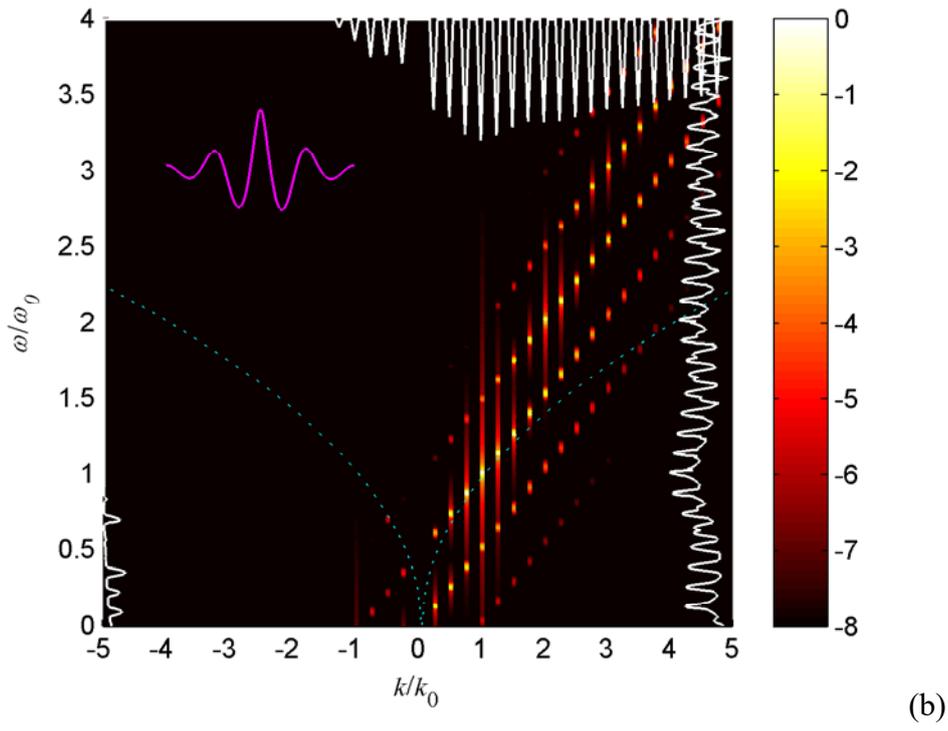

(b)

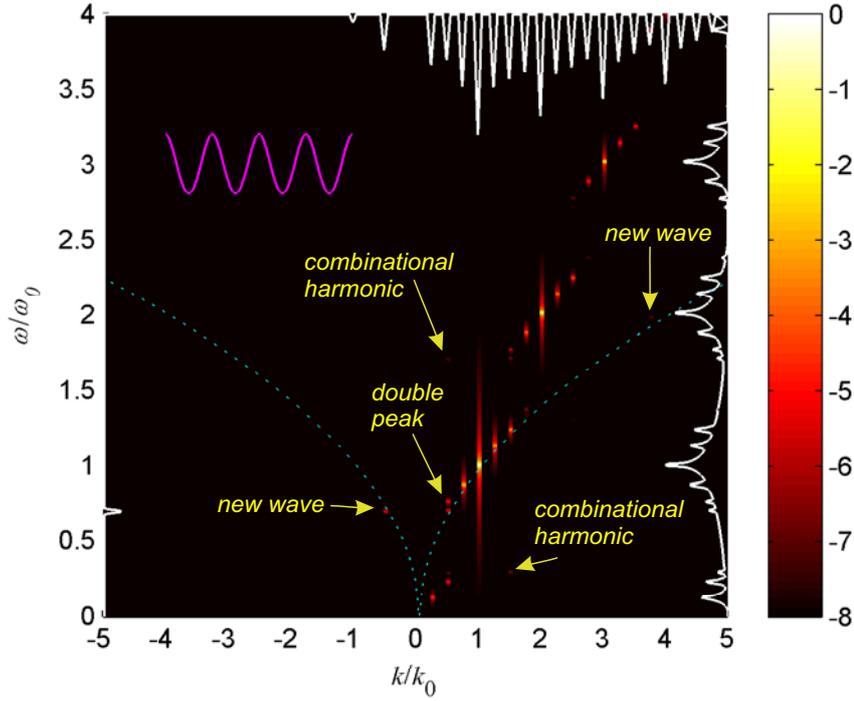

(c)

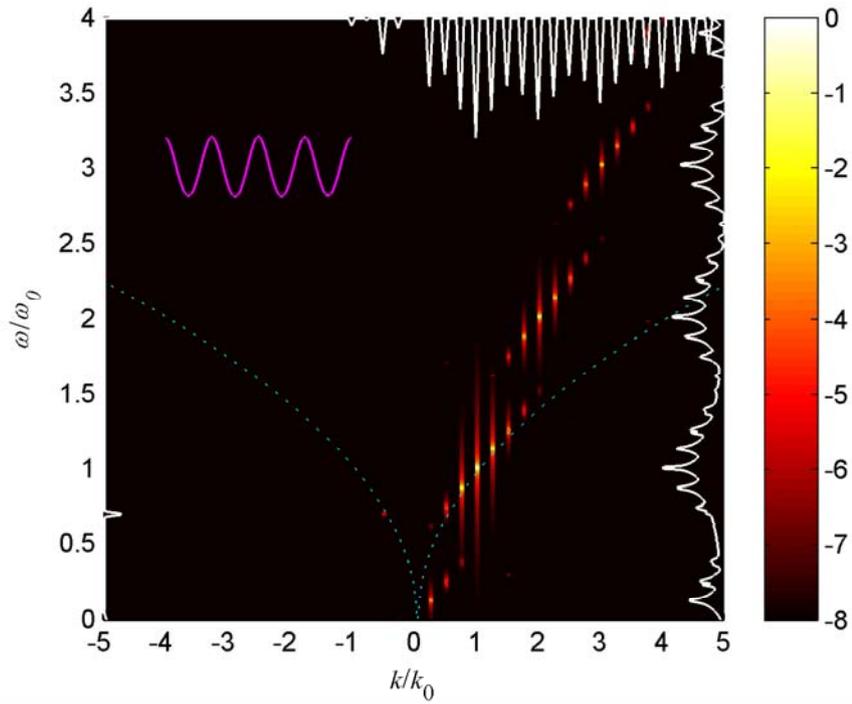

(d)

Fig. 9. The space-time Fourier transform amplitudes for the instants $t/T_0 = 347$ (a), 653 (b), 1000 (c) and 1200 (d) (see the selections in Fig. 7). The functions $\log_{10}S(\omega, k)$ are shown with the colour. The values $S(\omega, k)$ are normalized by the maximum value; the values below $10^{-8}$ are truncated. The white curves show the frequency Fourier amplitudes $S_{\omega-}$ and $S_{\omega+}$ (the left and right sides correspondingly), and the wavenumber Fourier amplitudes $S_k$ (the upper side) in the logarithmic coordinates; the values below $10^{-8}$ are truncated. The magenta curves at the left-upper quarters show the appearance of the surface displacement. The cyan broken lines plot the dispersive curves.

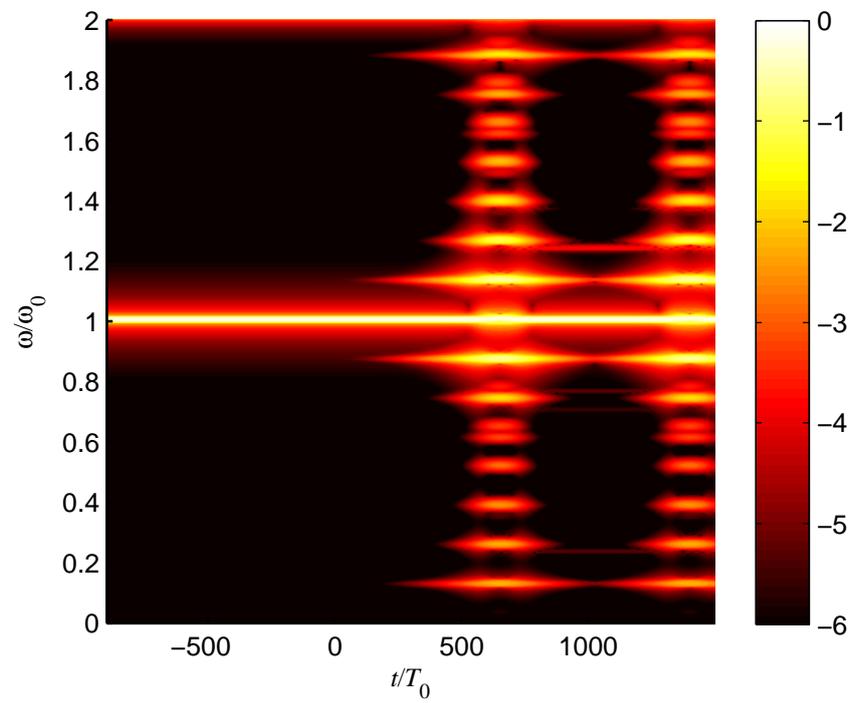

Fig. 10. The evolution of the windowed frequency Fourier transform $\log_{10} S_{\omega^+}$. The values of $S_{\omega^+}$ below $10^{-6}$ are truncated.